# Semiconductor Split-Ring Resonators for Thermally Tunable, Terahertz Metamaterials

**Jiaguang Han**[a]

Department of Physics, National University of Singapore, 2 Science Drive 3, Singapore 117542, Singapore

**Akhlesh Lakhtakia**[b]

NanoMM–Nanoengineered Metamaterials Group, Department of Engineering Science and Mechanics, Pennsylvania State University, University Park, PA 16802, USA

**Abstract**

As the variation of temperature alters the intrinsic carrier density in a semiconductor, numerical simulations indicate that the consequent variation of the relative permittivity in the terahertz regime provides a way to realize thermally tunable split-ring resonators. Electromagnetic metasurfaces and metamaterials that are thermally tunable in the terahertz regime can thus be implemented.

---

[a] E-mail Address: phyhanj@nus.edu.sg

[b] Corresponding Author. E-mail Address: akhlesh@psu.edu

# 1. Introduction

Metamaterials are composite materials designed to exhibit an optimized combination of two or more responses to specific excitation. Typically, this multifunctionality is uncommon if not impossible in nature, and emerges from the engineered cellular morphology of a metamaterial. In other words, a metamaterial is an engineered assemblage of different types of mutually inert or mutually reinforcing cells of sufficiently small dimensions [1].

Among electromagnetic metamaterials, those that refract negatively have drawn considerable attention during this decade [2–4]. A type of cell used commonly in negatively refracting electromagnetic metamaterials is the planar splint-ring resonator (SRR), which is a two-dimensional metallic structure printed on a circuit-board substrate. A planar SRR is a resonant structure. The scope of a metamaterial containing tunable planar SRRs as some of its constituent cells is obviously very high and encompasses, for example, spatial light modulators and tunable optical filters.

Tunability strategies examined thus far include electrical control [5–7], magnetostatic control [8,9], and optical pumping [10]. Whereas the planar SRRs remain metallic in the cited publications, tunability is introduced by ensuring that another component (e.g., the substrate on which the SRRs are printed or another type of inclusions) is made of an electro-optic material, liquid crystal, ferrite, etc.

In this communication, we theoretically demonstrate the feasibility of metamaterials that are tunable in the terahertz regime by change of temperature. These metamaterials would not comprise metallic SRRs, but semiconductor SRRs instead. In the terahertz regime, semiconductors emulate metals by having the real parts of their relative permittivities negative. Compared to metals, however, semiconductors have a significant advantage in that their relative permittivity can be modified by changing the temperature, thereby changing the carrier density and, therefore, the plasma frequency of the semiconductor.

**2. Preliminaries**

For the sake of definiteness, let us consider a planar SRR made of InSb printed on an isotropic quartz substrate. Fig. 1a shows the schematic of a single SRR with linear dimensions in the plane ranging from 2 to 36 μm, and of thickness 200 nm. The quartz substrate of relative permittivity 3.78 is 640-μm thick. The essence of the chosen type of metamaterial is a metasurface [11]: a square array of InSb SRRs with lattice parameter 50 μm, as shown in Fig. 1b. Computer simulations of the spectral response of this metasurface were performed using the commercial software CST Microwave Studio TM 2006B, which is a three-dimensional, full-wave solver employing the finite integration technique.

In the terahertz regime, the complex-valued relative permittivity of InSb is given by the simple Drude model [12]

$$\varepsilon(\omega) = \varepsilon_\infty - \omega_p^2 / (\omega^2 + i\gamma\omega), \qquad (1)$$

where $\omega$ is the angular frequency; $\varepsilon_\infty$ represents the high-frequency value; $\gamma$ is the damping constant; and the plasma frequency $\omega_p = \sqrt{Ne^2/\varepsilon_0 m^*}$ depends on the intrinsic carrier density $N$, the effective mass $m^*$ of free carriers, the electronic charge $e$, and the free-space permittivity $\varepsilon_0$. Compared to metals, the plasma frequency $\omega_p$ of InSb depends strongly on the temperature $T$. The intrinsic carrier density $N$ (in cm$^{-3}$) in InSb obeys the relationship [13]

$$N = 5.76 \times 10^{14} T^{3/2} \exp(-0.26/2k_B T), \qquad (2)$$

where $k_B$ is the Boltzmann constant and the temperature is in Kelvin. A variation in $N$ due to a variation in $T$ thus changes $\omega_p$. Consequently, in the far-infrared portion of the terahertz regime, $\varepsilon(\omega)$ of InSb is very sensitive to $T$. Hence, for metasurfaces and metamaterials comprising InSb SRRs, we can expect that temperature variations can cause substantial variations in the optical response characteristics.

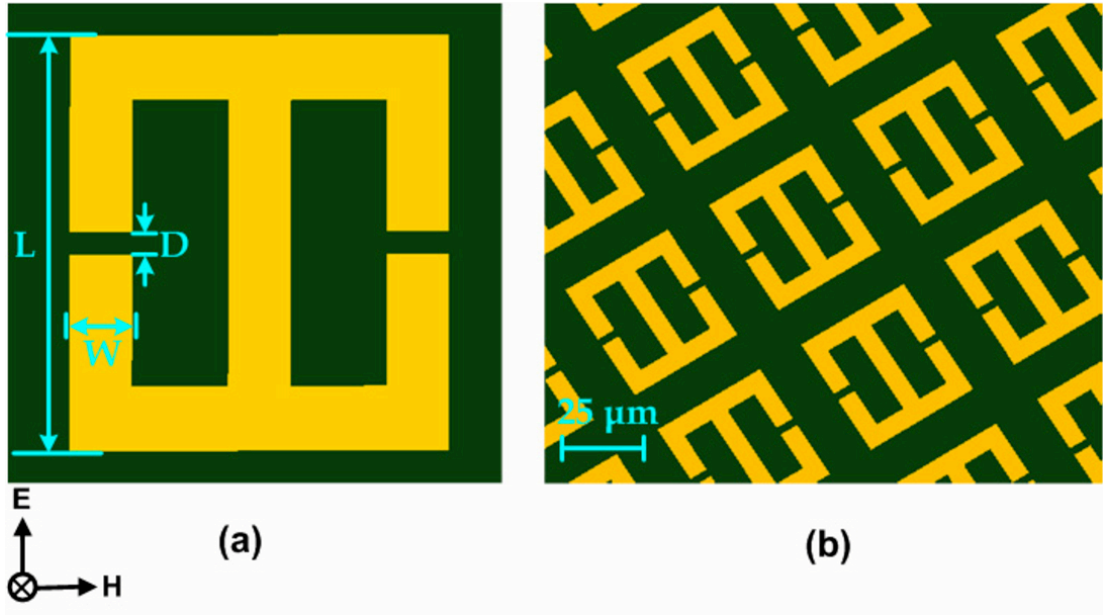

**FIG. 1**. (a) A single SRR with dimensions: L = 36 μm, D = 2 μm and W = 6 μm. (b) Schematic of a square array of SRRs printed on a quartz substrate with a period of 50 μm. The SRRs are made of InSb, whose relative permittivity is given by Eq. 1 with $\varepsilon_\infty$ =15.68, $m^* = 0.015 m_e$, $m_e = 9.1 \times 10^{-31}$ kg, and $\gamma/2\pi$ =0.05 THz [12,13]. The intrinsic free carrier concentration in InSb is [13]: $N = 5.76 \times 10^{14} T^{3/2} \exp(-0.26/2k_B T)$ cm$^{-3}$. The electric field of the normally incident plane wave is oriented perpendicular to the gaps in the SRRs.

Suppose the chosen metasurface lies parallel to the plane $z = 0$ of a Cartesian coordinate system such that the gaps in all SRRs are aligned parallel to the $x$ axis. Let the metasurface be illuminated by a plane wave whose wave vector is oriented parallel to the $z$ axis and whose electric field vector is aligned parallel to the $y$ axis. Using CST Microwave Studio TM 2600B, we computed the transmittance spectrum of the metasurface at frequencies below 1.0 THz at various temperatures.

## 3. Numerical Results and Discussion

Figure 2a shows the computed frequency-dependent transmission of the chosen metasurface at various temperatures. The transmission is the ratio of the magnitude of the electric field of the transmitted plane wave to that of the incident plane wave. No remarkable resonance is evident at temperatures below 200 K, and the transmission is almost independent of frequency at 1.0 THz and lower frequencies. If the temperature is increased to 220 K, a resonance is manifested as a small transmission dip around 0.09 THz. At 250 K, the resonance becomes more visible and occurs at 0.15 THz with a transmission minimum of 0.792. The resonance blueshifts to 0.28 THz with a transmission minimum of 0.549 at 300 K, and then further to 0.44 THz with a transmission minimum of 0.32 at 350 K. Clearly, the transmission characteristics of the metasurface change significantly with temperature.

Figure 2b shows a 3D profile of the resonant frequency corresponding with value of transmission dip as a function of temperature above 250 K, for which purpose detailed simulations were performed at a temperature interval of 10 K. As expected, the resonance of the metasurface can be significantly enhanced and blueshifted by raising the temperature.

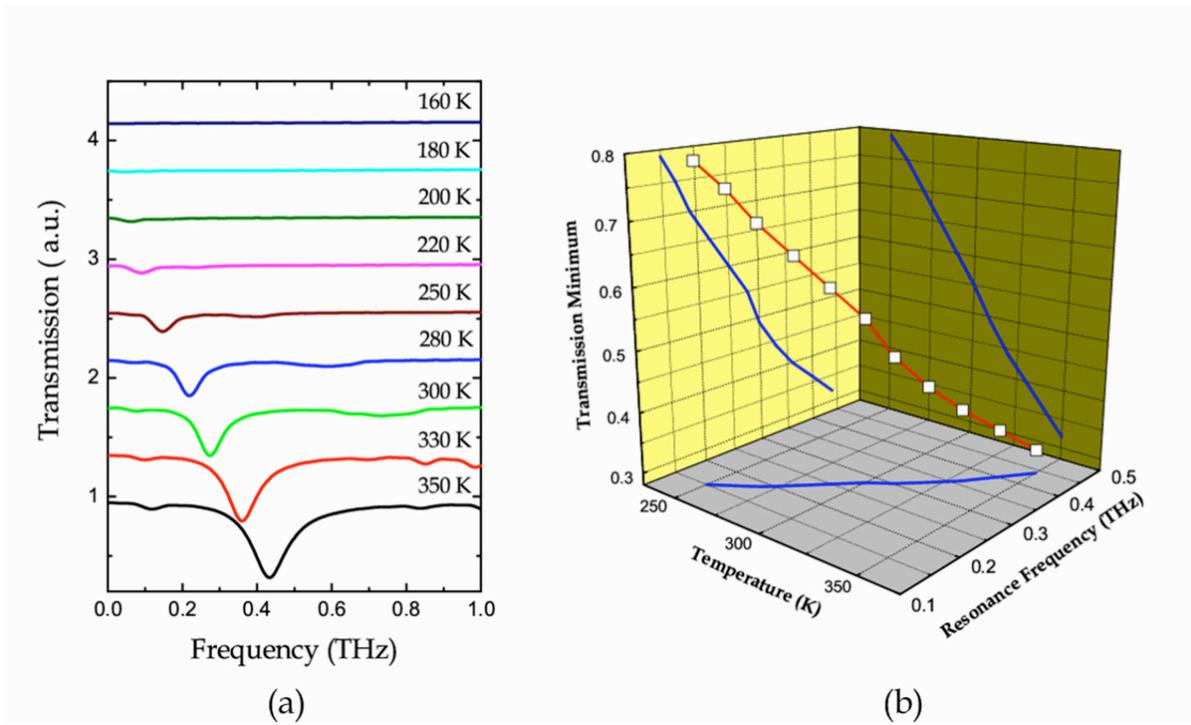

**FIG. 2**. (a) Transmission spectra of the chosen metasurface at different temperatures ranging from 160 to 350 K. (b) Variation of the remarkable transmission minimum and the resonance with temperature above 250 K.

The temperature-dependent resonant characteristics can be attributed directly to the increase in the density of free carriers by thermal excitation as the temperature increases. For instance, the intrinsic carrier density in InSb is $N \sim 0.94 \times 10^{14}$ cm$^{-3}$ at 160 K, but $N \sim 5.07 \times 10^{16}$ cm$^{-3}$ at 350 K; thus, the semiconductor displays more metallic features with increasing temperature.

This effect also can be seen from the variation of $\varepsilon(\omega)$ with $T$. The real part $\varepsilon_r(\omega)$ and the imaginary part $\varepsilon_i(\omega)$ of $\varepsilon(\omega)$ alter dramatically because of the variation of $N$ with $T$. At a specific value of $\omega$, $\varepsilon_r(\omega)$ reduces and $\varepsilon_i(\omega)$ increases as $T$ increases, as evinced by the plots in Fig. 3a. The more metallic behavior is also demonstrated by an increasing positive ratio $-\varepsilon_r(\omega)/\varepsilon_i(\omega)$ with temperature in Fig. 3b. Clearly, the InSb SRRs have almost insulating features at low temperatures but metallic features at high temperatures.

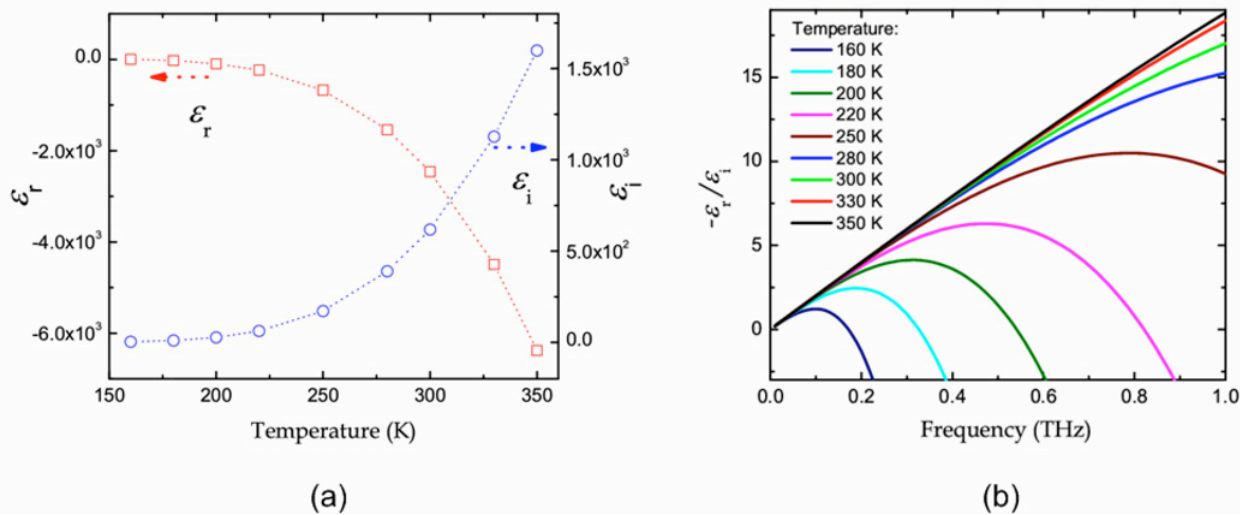

**FIG. 3**. (a) The real part $\varepsilon_r(\omega)$ (open squares) and the imaginary part $\varepsilon_i(\omega)$ (open circles) of the relative permittivity of InSb at 0.2 THz as functions of temperature. (b) Frequency-dependent ratio $-\varepsilon_r(\omega)/\varepsilon_i(\omega)$ of InSb in the terahertz regime at different temperatures.

The resonance frequency turns out to be almost linearly related to $-\varepsilon_r/\varepsilon_i$, as shown in Fig. 4a. The enhancement of the resonance feature by increase of temperature is increased is further evident from the consequent increase of the Q factor, as presented in Fig. 4b. This increase of $Q$ with $T$ is ascribed to the better metallic characteristics of InSb at a higher temperature, and this conclusion is consistent with the recent experimental observation that an SRR made of a better metal has a more pronounced resonance [14].

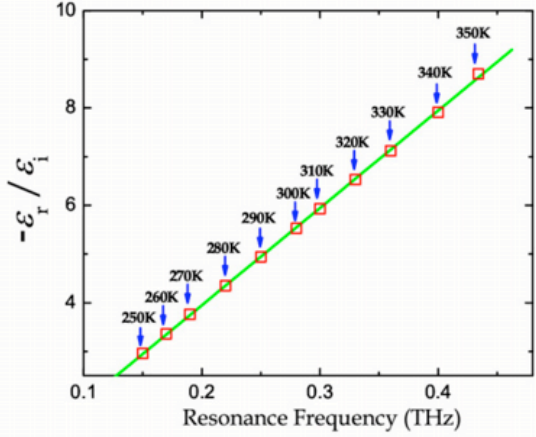 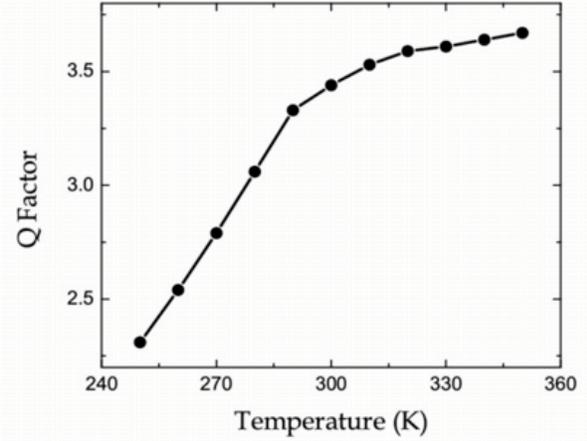

**FIG. 4**. (a) The ratio $-\varepsilon_r(\omega)/\varepsilon_i(\omega)$ at the resonance frequency at different temperatures. The solid line indicates a good linear fit. (b) The quality factor $Q$ of the transmission dip at different temperatures.

## 4. Concluding Remarks

So we have shown that as a variation of temperature alters the intrinsic carrier density in a semiconductor, the consequent variation of the relative permittivity in the terahertz regime provides a way to make thermally tunable split-ring resonators and even other types of cells for use in assembling electromagnetic metamaterials. These meteametarials would therefore be thermally tunable in the terahertz regime. Furthermore, any other mechanism for varying the intrinsic carrier density would endow tunability to electromagnetic metamaterials comprising semiconducting split-ring resonators, etc.

**Acknowledgements.** JH acknowledges fruitful discussions with Prof. Takeda and Dr. Miyamaru (Shinshu University, Japan) and financial support from the MOE Academic Research Fund of Singapore.